\documentclass[aps,pre,twocolumn,floatfix,superscriptaddress,showpacs,showkeys]{revtex4-1}
\usepackage{epsf,amsmath,amssymb,verbatim,color,multirow,pifont}
\usepackage{graphicx}
\usepackage{braket}
\usepackage{cleveref}
\usepackage{newfloat,algcompatible}
\usepackage{etoolbox}
\usepackage{tabularx}
\usepackage[normalem]{ulem}

\crefname{equation}{Eq.}{Eqs.}
\crefname{section}{Sec.}{Secs.}
\crefname{figure}{Fig.}{Figs.}
\crefname{table}{TABLE}{Figs.}

\begin{document}
\title{Forest Fire Model as a Supercritical Dynamic Model in Financial Systems}
\author{Deokjae Lee}
\affiliation{Center for Complex Systems Studies and CTP,  Department of Physics and Astronomy, Seoul National University, Seoul 151-747, Korea}
\author{Jae-Young Kim}
\affiliation{Department of Economics, Seoul National University, Seoul 151-747, Korea}
\author{Jeho Lee}
\email{jeho0405@snu.ac.kr}
\affiliation{Graduate School of Business, Seoul National University, Seoul 151-747, Korea}
\author{B. Kahng}
\email{bkahng@snu.ac.kr}
\affiliation{Center for Complex Systems Studies and CTP, Department of Physics and Astronomy, Seoul National University, Seoul 151-747, Korea}
\date{\today}

\begin{abstract}
Recently, large-scale cascading failures in complex systems have garnered substantial attention. Such extreme events have been treated as an integral part of the self-organized criticality (SOC). Recent empirical work has suggested that some extreme events systematically deviate from the SOC paradigm, requiring a different theoretical framework. We shed additional theoretical light on this possibility by studying financial crisis. We build our model of financial crisis on the well-known forest fire model in scale-free networks. Our analysis shows a non-trivial scaling feature indicating supercritical behavior, which is independent of system size. Extreme events in the supercritical state result from bursting of a fat bubble, seeds of which are sown by a protracted period of a benign financial environment with few shocks. Our findings suggest that policymakers can control the magnitude of financial meltdowns by keeping the economy operating within reasonable duration of a benign environment.
\end{abstract}


\maketitle

Large-scale cascading failures have garnered attention in many complex systems, such as power grids and communication networks ~\cite{buldyrev_catastrophic_2010,jacobson_congestion_1988,hancock_cell_2010,lee_branching_2012}, 
because once they happen, their impact can be unexpectedly catastrophic. A case in point is the crippling blow to the world economy preceded by the failure of an investment bank, Lehman Brothers, and the subsequent financial meltdown with the evaporation of more than \$10 trillion from the global equity market \cite{stiglitz_freefall:_2010}. In the past, such an extreme event was treated as an integral part of the self-organized criticality ~\cite{bak_self-organized_1987,scheinkman_self-organized_1994,jensen_self-organized_1998}, which is characterized by a power-law distribution. Partly due to the scarcity of extreme events, few suspected the possibility that some of them could systematically deviate from a power-law distribution. Recently, however, researchers have begun to consider extreme events as supercritical phenomena, characterizing extreme events as distinguishable by their sizes from the rest of the statistical population \cite{sornette_why_2004,sornette_dragon-kings:_2012}. The objective of our work is to shed additional light on such supercritical behavior by studying financial meltdown.

We build our model of financial crisis on the existing forest fire (FF) model introduced by Drossel and Schwabl ~\cite{drossel_self-organized_1992,malamud_forest_1998,jagla_forest-fire_2013}, because it captures two essential features in financial meltdown. First, its non-conservative ingredient naturally mimics financial meltdown, where asset prices tend not to be conserved. When an asset market collapses, traders have difficulty pricing assets, as was the case in the collapse of the mortgage-backed securities market on the eve of the 2008 Financial Crisis. The assets that were previously considered liquid become illiquid, causing chronic problems for banks with speculative bets on these assets. The upshot is that an important quantity, the value of assets, will not be conserved over time. Second, there exists a separation of two time scales. It takes a long time for banks to build up a fat bubble, which is represented by a percolation cluster consisting of counterparties of vulnerable banks that make speculative bets on risky assets. In contrast, the meltdown of this cluster takes place very quickly as trees burn up in a short time.

Here, we model the FF dynamics in scale-free networks, which are employed to capture entangled counterparty relationships among banks worldwide. For example, on the eve of the 2008 financial crisis, Lehman alone was counterparty to almost a million derivatives contracts and a huge borrower in the repo market, and its zillions of derivative and repo contracts connected the bank to numerous counterparties all over the world ~\cite{blinder_after_2013}. Our analysis shows a non-trivial scaling feature indicating supercritical behavior, which is independent of system size. Prior research on the FF model did not detect this supercritical behavior ~\cite{drossel_self-organized_1992,schenk_finite-size_2000,grassberger_critical_2002,pruessner_broken_2002}. We are able to detect it because it becomes more pronounced and conspicuous in scale-free networks, where the percolation threshold vanishes when the degree exponent is between two and three. 

\begin{figure*}[t]
\includegraphics[width=.9\textwidth]{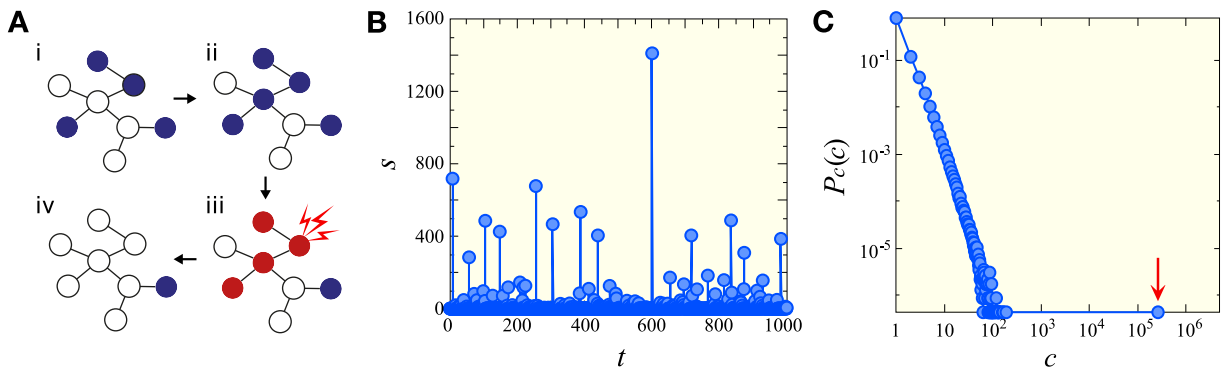}
\caption{(Color online) (A) A schematic illustration of the model. 
i) Empty (filled) nodes represent healthy (vulnerable) banks. 
ii) A randomly chosen healthy bank becomes vulnerable by taking excessive risks, 
and a cluster develops composed of four connected vulnerable banks. 
iii) One of the banks in the cluster is exposed to a random shock (represented as lightning), 
the exposed bank fails, and a cascade of bank failures is triggered throughout the entire cluster. 
iv) Those failed nodes become healthy. 
The number of failed banks in iii) is avalanche size. 
(B) A typical sequence of the avalanche sizes in a network with $10^5$ banks. Most avalanches are small,  
but a large-scale financial meltdown does occur as shown at time step $600$. 
(C) The cluster size distribution of inter-connected vulnerable banks at an onset of a large-scale financial meltdown. 
There exists a giant cluster (arrow), where complex transactional relationships among banks in vulnerable state will serve as a channel for financial crisis contagion.
}
\label{Fig-SchematicIllustration}
\end{figure*}

{\it Model:} Building on~\cite{drossel_self-organized_1992}, we model the contagion of 
financial crisis through an inter-banking network of size $N$, which is represented by a scale-free network with the degree distribution $P_d(k) \sim k^{-\gamma}$ ~\cite{lu_complex_2006,cho_percolation_2009}.
It is known to be ubiquitous, and empirical research suggests that an inter-banking network can be approximated by a scale-free network ~\cite{boss_network_2004,soramaki_topology_2007}. 
In the inter-banking network, each node represents a bank or bank-like firm, whereas a link between two nodes represents a counterparty relationship.
A bank may lend money to its counterpart bank or invest in its financial products or assets. When one bank defaults on some debt, this event can leave its counterpart creditors or investors dangerously short of funds. To shed some meaningful light on the dynamics of such a complex system, our model focuses only on cascading bank failures in the inter-banking network. Defaults of non-financial firms or individuals are treated as external shocks to the system. 

The dynamics of the FF model in the inter-banking network is defined as follows: Each node can be in one of the two states: vulnerable or healthy, which corresponds to a tree-occupied state or an empty state. In the vulnerable state, the node has insufficient cash reserves and is susceptible to financial shock. In the healthy state, the bank has enough cash or liquid assets on hand to meet depositors' (or creditors') demands, and is resilient to financial shocks. Initially all nodes are healthy, and the following steps are repeated: i) a randomly chosen node becomes vulnerable; ii) a randomly chosen node experiences a shock with a probability of $1/\theta$. If the chosen node is vulnerable, the whole cluster of vulnerable nodes containing the chosen node fails, and all the failed nodes become healthy. This approach to modeling of financial contagion differs from typical epidemic models, where healthy individuals are susceptible to infection from infected individuals. 
Actually such contact processes are supposed to exist but are ignored in the FF model 
because their time scale is too short compared with that of growing trees. 
We call the number of nodes in the failed cluster the avalanche size.
The probability distribution of avalanche sizes, which is denoted as $P_s(s)$, is our primary interest (Fig.~\ref{Fig-SchematicIllustration} A,B).

{\it Implication to financial systems:} The parameter $\theta$ controls the average duration between two successive external shocks (two successive instances of lightening in the context of forest fire), which may be interpreted as the availability of liquidity in a financial system. In the model, the extreme events result from bursting of bubbles, seeds of which are sown by economic 
expansion with few shocks for a long period, which corresponds to the case when $\theta$ is large. 
That is, as banks do not experience defaults on their loans, more and more banks get involved in transactions of risky assets with many other counterparties, building up an extremely fat bubble.
Historically, the fragility of the financial system has been increased by long periods of easy access to money,
during which defaults on loans were infrequent \cite{roubini_crisis_2011,krugman_end_2013}.

After the expansion with easy money, the moment arrives for a dramatic reversal of the expansion$-$this is now known as a Minsky moment in the financial community \cite{minsky_can_1984,minsky_stabilizing_2008}. 
Usually an external shock, such as sudden increases in interest rates, acts as a wakeup call of a financial meltdown. Assets that were previously considered liquid become illiquid and values of risky assets are heavily devaluated. Banks with imprudent practices can no longer borrow money from the inter-banking money market at a reasonable cost and fail.
The devaluation and the propagation of failures induce each other amplifying the meltdown \cite{minsky_can_1984,minsky_stabilizing_2008,krugman_return_2009,roubini_crisis_2011}.
On the other hand, in the FF model, there are no locally conserved ``carriers'' of vulnerability such as sand grains in the sandpile model \cite{bak_self-organized_1987}. The failure of a node simply causes failures of all vulnerable nodes connected to the failed node. This is a simplified version of the real situation in which vulnerability is amplified by the collapse of asset markets and the subsequent evaporation of liquidity. 
The non-conservative nature of the FF model is an essential ingredient for the supercritical behavior because conventional conservative avalanche models such as sandpile model do not exhibit supercriticality in regular lattices or scale-free networks \cite{goh_sandpile_2003}.

In the FF model, the separation of two time scales, the periods of expansion and meltdown, seems to be reasonable for modeling financial crisis. After the Minsky moment, a failure of one vulnerable bank tends to trigger a financial meltdown. Since the timescale for such a meltdown in reality is much smaller than that for expansion \cite{stiglitz_freefall:_2010}, we describe a financial meltdown as a series of bank failures occurring in one time step. 

In the next time step, the failed nodes become healthy again.
An interpretation of this rule is that the failed banks are refinanced through government bailouts or acquisitions by other actors. 
In reality, failed banks may also be dissolved, or new banks may enter the system, and the bank network evolves. 
However, after a transient period of the evolution which is rather short compared with the interval between two successive financial crises, the network should be still scale-free and have similar statistical properties with the previous one.
Thus, statistical properties of the FF model on such dynamic networks can be obtained by repeated simulations in an ensemble of scale-free networks.

\begin{figure}[t]
\includegraphics[width=.9\linewidth]{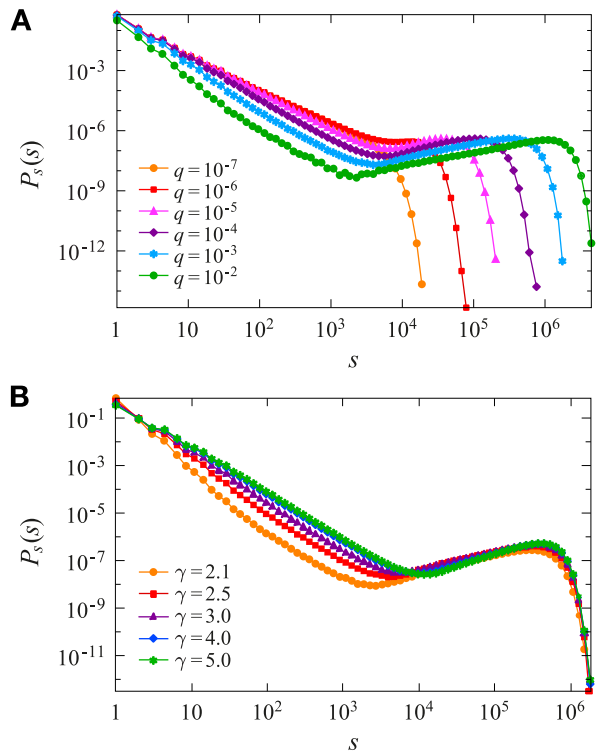}
\caption{(Color online) 
(A) Probability distributions of avalanche sizes for various $q$. 
The tails of the distributions systematically deviate from a power-law behavior. 
(B) Probability distributions of avalanche sizes for various degree exponent values $\gamma$. 
Simulations were run on scale-free networks with exponent $\gamma=2.5$ for (A), containing $N=10^7$ nodes and $L=10N$ links for (A) and (B).
}
\label{Fig-MainResult}
\end{figure}

\begin{figure}[t]
\includegraphics[width=.92\linewidth]{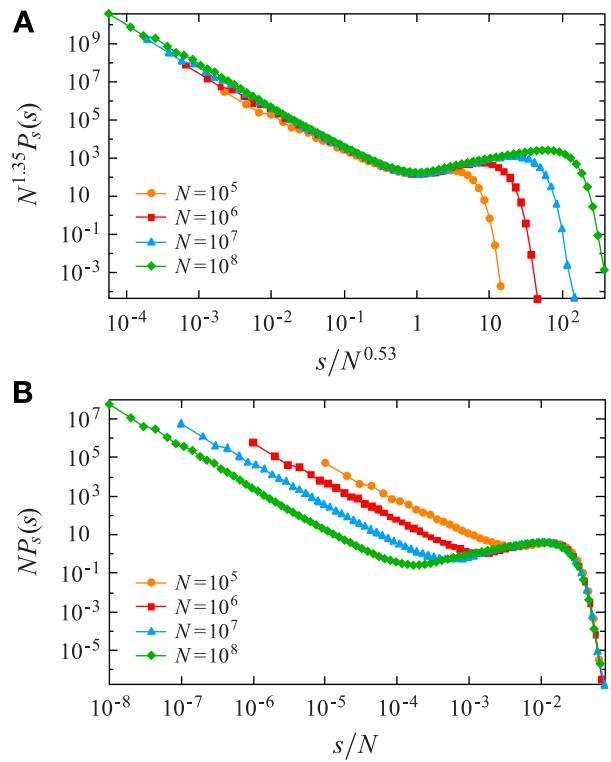}
\caption{(Color online) 
(A) Plot of data collapse in small-size avalanche region and (B) intermediate-size avalanche region for different system sizes $N$. 
The parameter $q$ is fixed as $10^{-4}$. The underlying networks are scale-free networks with degree exponent $\gamma=2.5$.
(A) indicates the crossover point $s_{c1}$ between the first and second region scales as $N^{0.53}$ 
and (B) indicates the other crossover point $s_{c2}$ between the second region and third region scales as $N$.
}
\label{Fig-Scaling}
\end{figure}

{\it Avalanche size distribution:} The simulation results of $P_s(s)$ for various $q \equiv \theta/N$ are shown in Fig.~\ref{Fig-MainResult}A. 
They show that a parameter $q$, the availability of liquidity relative to the system size, positively affects the magnitude of the large-scale financial meltdown. 
When $q$ is sufficiently small, the size distribution decays in a power law-like manner. 
In this case, external shocks are frequent, small-scale avalanches are more likely, and large-scale avalanches are less likely. 
In contrast, when $q$ is large, the distribution exhibits a supercritical behavior and can be characterized in three distinct regions: 
i) in the first region, the size distribution decays at a rate close to a power-law; 
ii) in the second region, a bump exists whose pattern can be described by an increasing power-law function, 
called the supercritical region; 
iii) in the third region, the distribution tails off sharply. 

In the model, a protracted period without shock allows the development of bubbles, 
which are represented by a giant cluster of complex transactional relationships among vulnerable banks. 
This is equivalent to the giant cluster in a percolation theory \cite{cohen_percolation_2002} 
(Fig.~\ref{Fig-SchematicIllustration}C).
The failure of one bank in a giant cluster causes the failure of the whole cluster. It is a large-scale financial meltdown in the model.
The size distribution in the first region is due to the failures of banks in finite-sized clusters as well as giant clusters, 
whereas the distribution in the second region stems from the failures of banks in giant clusters only. 

We examine the sensitivity of our key findings to a change in the degree exponent $\gamma$, which controls the degrees of mega banks. 
For all levels of $\gamma$ from $2.1$ to $5$, supercritical behavior is apparent (Fig.~\ref{Fig-MainResult}B).
We also run simulations on regular lattics, in which mega banks are outright absent. 
Supercritical behavior also appears when $q$ is sufficiently large, which will be shown later.
This result shows that the absence of mega banks does not eliminate the possibility of a supercritical financial meltdown completely if $q$ is large.
Indeed, history suggests that large-scale financial meltdowns did occur in pre-modern eras prior to the evolution of modern mega banks 
\cite{roubini_crisis_2011,mackay_memoirs_2009,kindleberger_manias_2005}.

{\it Finite-size scaling of supercritical behavior}: 
A bump in an avalanche size distribution is found in many systems.
Such bumps are usually believed to be a finite-size effect that vanishes in the thermodynamic limit, manifesting critical behavior.
However, the bump in our avalanche size distribution is qualitatively different in that it sustains in the thermodynamic limit, implying genuine supercritical behavior.
Furthermore, the bump exhibits the increasing power-law behavior, which has not been observed in other avalanche dynamics to our knowledge.
Here we systematically analyze these observations based on a finite-size scaling analysis of the avalanche size distribution.

We first denote the crossover point between the first and second regions as $s_{c1}$. 
In Fig.~\ref{Fig-Scaling}, we show that the aforementioned behavior is observed in systems of different sizes, 
but the crossover point depends on the system size in a power-law manner (i.e., $s_{c1} \sim N^\mu$). 
Based on this result, we make the usual scaling ansatz:
\begin{equation}
P_s^{(<)}(s) = c_1 G_<(s/s_{c1}).
\end{equation}
The scaling function $G_<(x)$ behaves as $G_<(x) \sim x^{-\tau}$ for $x < 1$. 
To eliminate the size dependency, $c_1$ is determined as $c_1 \sim N^{-\mu \tau}$.

The crossover point between the second and the third regions is denoted as $s_{c2}$. 
To characterize the scaling behavior in the bump pattern for different system sizes, 
we introduce another scaling hypothesis:
\begin{equation}
P_s^{(>)}(s) = c_2 G_>(s/s_{c2})
\end{equation}
where $s_{c2} \sim N$, because $s_{c2}$ represents a massive-scale avalanche comparable to the system size in order of magnitude (Fig.~\ref{Fig-Scaling}B). 
Then the scaling function behaves as $G_>(x) \sim x^\zeta$ for $x < 1$ and sharply decays for $x > 1$. 
Because the two avalanche size distribution functions are continuous at $s_{c1}$,
the coefficient $c_2$ must depend on the system size as $c_2 \sim N^{\zeta - \mu(\tau +\zeta)}$.

We numerically confirm the scaling behavior using the data collapse procedure. 
The scaling hypothesis implies that curves $N^{-\mu\tau} P_s^{(<)}(s/N^\mu)$ for different $N$ should collapse into the same curve in the first region, 
and that the collapsing part extends as $N$ grows.
Similarly, the curves $N^{-\zeta + \mu(\tau + \zeta)} P_s^{(>)}(s/N)$ for different $N$ should collapse into the same curve in the second region. This data collapse is well established by the choice of $\tau=2.55$, $\mu=0.53$, and $\zeta=0.75$ for networks with $\gamma=2.5$ (Fig.~\ref{Fig-Scaling}). 
The chosen exponents do not depend on $q$, but on $\gamma$ (Table.~\ref{Table-Exponents}). 
Thus, the scaling behavior is independent of the parameter $q$. 
It depends only on the topology of the underlying network.

A scaling relation between the exponents $\tau$, $\mu$, and $\zeta$ is obtained by considering the average size of avalanches $\left< s \right>$. 
For a duration $\Delta t$, the average number of avalanches is given by $\rho \Delta t / q N$, 
where $\rho$ is the density of the vulnerable banks in the steady state. 
Then, the average number of failed banks in the duration is $\left< s \right> \rho \Delta t / q N$. 
This must be equal to the average number of banks $(1-\rho) \Delta t$ that become vulnerable in the duration 
because the number of vulnerable banks in the system is steady on average. 
Thus, we obtain $\left< s \right> = q N (1 - \rho) / \rho$ \cite{drossel_self-organized_1992}. 
On the other hand, we have 
$\left< s \right> = \int_1^{s_{c1}} s P_s^{(<)} (s) ds + \int_{s_{c1}}^{\infty} s P_s^{(>)}(s) ds \sim N^{\zeta - \mu(\tau + \zeta) + 2}$. 
Thus, the relation 
$\mu(\tau + \zeta) - \zeta = 1$
is obtained. Our measurement of the exponents in the simulations also fits the relation well. 
{\it This relation implies that the second region is sustained in the thermodynamic limit 
because the probability of the avalanches in the region is constant as}
\begin{equation}  
\int_{s_{c1}}^{s_{c2}} P_s^{(>)}(s) ds \sim N^{\zeta - \mu(\tau + \zeta) + 1} \sim {\rm const}.
\end{equation}

\begin{figure}[t]
\includegraphics[width=.9\linewidth]{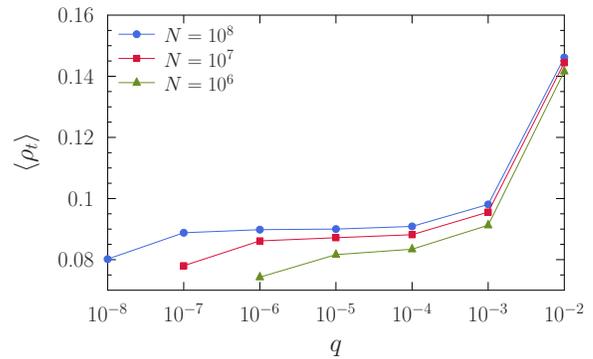}
\caption{(Color online) 
Average density of vulnerable nodes as a function of $q$. The density is measured just before each shock. 
Degree exponent of the underlying network is taken as $\gamma=2.5$.
The scaling behavior is evident in the range of $q$ where the plateau is observed in the average density. 
}
\label{Fig-Density}
\end{figure}

\begin{figure}[t]
\includegraphics[width=.9\linewidth]{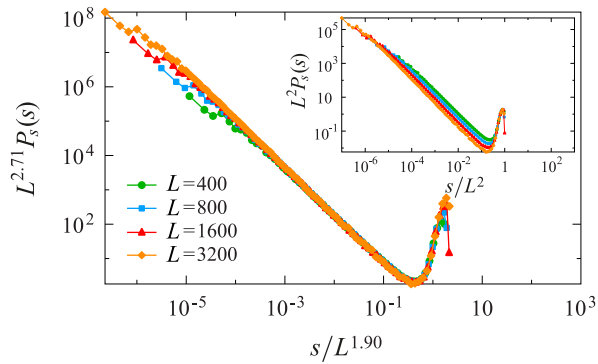}
\caption{(Color online) 
Data collapse of the avalanche size distributions on two dimensional square lattices of different sizes for 
a fixed $q=0.7$. The linear size of a lattice is denoted as $L$.
The collapse of the decreasing part indicates that the crossover point scales as $L^{1.9}$. 
The inset shows the cutoff point of the increasing part scales along the order of the system size $L^2$.}
\label{Fig-2DLattice}
\end{figure}

The range of $q$ in which the scaling behavior holds can be estimated by 
the average density of vulnerable nodes measured just before each shock. 
We find that there exists a range of $q$ in which the average density measured just before each shock is independent of $q$.
The scaling behavior holds in the range, whereas deviation is observed outside the range (Fig.~\ref{Fig-Density}).
A finite upper bound $q_0$ ($\approx 10^{-4}$) of the range exists, but the lower bound seems to vanish as the system size increases.
For any $0 < q < q_0$, the scaling behavior holds if the system size is large enough.
The vanishing lower bound is supported by the fact that the percolation threshold vanishes in scale-free networks with $2 < \gamma <3$ \cite{cohen_percolation_2002}.
The density of vulnerable nodes is always higher than the percolation threshold, implying a supercritical 
giant cluster exists in the thermodynamic limit.

We have two divergent size scales $s_{c1}$ and $s_{c2}$ in the model.
The scaling of $s_{c2}$ of the order of $N$ is clear, but scaling of the crossover point $s_{c1}$ is not trivial.
To understand the characteristics of the exponent $\mu$, we simulate the model on the two dimensional square lattice. 
When $q$ is sufficiently large, a bump again appears in the tail of the distribution (Fig.~\ref{Fig-2DLattice}).
We find that the crossover points scale as $s_{c1} \sim L^{1.9}$ and $s_{c2} \sim L^2$, where $L$ is the linear size of the system and $N=L^2$.
The value $1.9$ of the exponent $s_{c1}$ is close to the fractal dimension of the percolating cluster around the percolation threshold on the lattice. 
The scaling is expected because $s_{c1}$ is the starting point of the bump and thus represents the typical size of small giant clusters.
The existence of the bump and the cutoff of the order of $N$ are also expected if the density of vulnerable nodes is maintained to be larger than the percolation threshold by sufficiently large $q$.
However, the bump is too narrow to exhibit any power-law behavior. Therefore, the arguments we made for the FF model on scale-free networks will not be directly applicable to the bump in the lattice.
We remark that in a previous study \cite{schenk_finite-size_2000}, a similar distribution was observed and was interpretted as a violation of simple finite-size scaling ansatz, 
but no scaling analysis for $s_{c1}$ was provided.

\begin{table}[b]
\centering{
\begin{tabularx}{0.9\linewidth}{|X|X|X|X|X|X|}
\hline
\centering{$\gamma$} & 2.1 & 2.5  & 3.0  & 4.0  & 5.0  \\ \hline
\centering{$\tau$}   & 2.8 & 2.55 & 2.4  & 2.0  & 1.98 \\ \hline
\centering{$\mu$}    & 0.5 & 0.53 & 0.55 & 0.65 & 0.67 \\ \hline
\centering{$\zeta$}  & 0.8 & 0.75 & 0.7  & 0.9  & 1.0  \\
\hline
\end{tabularx}
\caption{
Dependency of the scaling exponents on the degree exponent $\gamma$ of the underlying scale-free networks.
}
}
\label{Table-Exponents}
\end{table}

{\it Discussion:}
We have studied the FF model on scale-free networks and derived a scaling relation for the avalanche sizes in the supercritical region, which implies that the supercritical dynamics can occur generically, independent of system size. In particular, the dynamics of our model generate not only small-scale bank failures but also  extremely large-scale financial meltdowns. The dynamics are shaped by the formation of a bubble, which is represented by a cluster of counterparty relationships among vulnerable banks that make speculative bets on risky assets. 

The size of a bubble is controlled by the duration in which the system is not exposed to external shocks (e.g., no lightening in the context of forest fire). When the duration is short, small-scale bank failures are more likely, but the possibility for an extreme financial meltdown is reduced. When the duration is long enough, however, small-scale bank failures become less frequent, as is usually the case in an era of easy access to money. History, however, suggests that a protracted era of easy money promotes imprudent banking practices and development of speculative bubbles \cite{roubini_crisis_2011,krugman_end_2013,shiller_irrational_2005}. In our model, the system evolves to a supercritical state in this munificent environment, increasing the likelihood of development of an unusually large cluster of counterparty relationships among vulnerable banks with speculative bets on risky assets. This cluster is equivalent to a supercritical percolation cluster in the context of forest fire. When one bank in this cluster fails, other counterparties in the cluster are affected, and cascading bank failures occur.

This is reminiscent of the financial meltdown triggered by the demise of Lehman Brothers, which was acting as the prime broker for many hedge funds in executing trades, holding collateral, receiving, and disbursing monies \cite{blinder_after_2013}. Lehman's failure immediately wiped out plenty of hedge funds. The bankruptcy of Lehman's European subsidiary alone froze \$40 billion in clients’ funds \cite{buchanan_forecast_2014}. Furthermore, its biggest counterparties, such as Bank of America, Citigroup, and Deutsche Bank, were critically affected and eventually bailed out by governments. Then, a credit crunch hammered banking systems globally, and the shutdown of some asset markets made it difficult to conserve the value of an asset. It became blatantly obvious that this non-conservative nature is one of the essential features of financial crisis. Our findings highlight the importance of policy interventions in keeping the economy operating within reasonable duration of easy money regime, which seems to be one of the root causes of large-scale financial meltdowns.

\begin{acknowledgments}
This work was supported by the SNU research grant in form of brain fusion project and the NRF in Korea with grant Nos. 2010-0015066 and 2014-069005.
\end{acknowledgments}

%\bibliography{References}{}
%\bibliographystyle{apsrev4-1}


\end{document}